\documentclass[11pt,titlepage]{article}

\usepackage{setspace}
\usepackage{amsmath,amssymb,lastpage,bm,textcomp}
\usepackage{graphicx}
\usepackage[round,numbers,sort&compress]{natbib}
\usepackage{float}
\usepackage{authblk}
\usepackage[usenames,dvipsnames]{color}

\newcommand{\erfc}{\operatorname{erfc}}      

\def\tmin{t_\text{min}}
\def\tmax{t_\text{max}}

\title{Force-clamp analysis techniques reveal stretched exponential unfolding kinetics in ubiquitin}

\author[*]{Herbert Lannon}
\author[+]{Eric Vanden-Eijnden}
\author[*]{J. Bruji\'c\thanks{Corresponding author. 608 Meyer Hall, 4 Washington Place, New York, NY, 10003, USA, Tel:~(212)998-3586, Email: jb2929@nyu.edu}}
\affil[*]{Department of Physics and Center for Soft Matter Research, New York University, 4 Washington Place, New York, NY, 10003, USA}
\affil[+]{Courant Institute of Mathematical Sciences, New York University, New York, NY, 10012}

\date{}

\begin{document}

\maketitle
\abstract{ 
Force-clamp spectroscopy reveals the unfolding and disulfide bond rupture times of single protein molecules as a function of the stretching force, point mutations and solvent conditions. The statistics of these times reveal whether the protein domains are independent of one another, the mechanical hierarchy in the polyprotein chain, and the functional form of the probability distribution from which they originate. It is therefore important to use robust statistical tests to decipher the correct theoretical model underlying the process. Here we develop multiple techniques to compare the well-established experimental data set on ubiquitin with existing theoretical models as a case study. We show that robustness against filtering, agreement with a maximum likelihood function that takes into account experimental artifacts, the Kuiper statistic test and alignment with synthetic data all identify the Weibull or stretched exponential distribution as the best fitting model. Our results are inconsistent with recently proposed models of Gaussian disorder in the energy landscape or noise in the applied force as explanations for the observed non-exponential kinetics. Since the physical model in the fit affects the characteristic unfolding time, these results have important implications on our understanding of the biological function of proteins.  

}

\clearpage

\section*{Introduction}

Force-clamp spectroscopy using the atomic force microscope (AFM) has proven to be a useful tool for following the unfolding trajectories of single polyprotein molecules~\cite{Oberhauser2001, Fernandez2004, YiCao2008, Liu2009, Perez-Jimenez2006, Bullard2006}. Previous studies have investigated the effect of the applied force~\cite{Schlierf2004, Brujic2006, Garcia-Manyes2009, Kuo2010, Liu2009}, length of the polyprotein chain~\cite{Brujic2007, Garcia-Manyes2007} and order statistics~\cite{YiCao2008} on the unfolding kinetics of mechanically stable proteins. The simplest free energy landscape model for mechanical unfolding is a two-state reaction process over a single transition state barrier, which is tilted by the work done on the molecule~\cite{Bell1978}. In such a reaction driven by simple diffusion, the probability distribution of the measured dwell times at a given force is exponential with a rate of decay that is governed by the barrier height. Moreover, the unfolding rate is exponentially dependent on the applied force. The majority of previous studies have interpreted their data using this two-state model to determine the height of the energy barrier and the distance to the transition state.

Apart from the two-state fitting of the unfolding kinetics of ubiquitin~\cite{Schlierf2004}, more recent work has shown that a larger statistical pool of dwell times at a given force reveals important deviations from exponential kinetics and requires more sophisticated modeling. Surprisingly, these deviations have led to three alternative models with different physical interpretations for the unfolding of ubiquitin pulled under the same experimental conditions. The first physical interpretation considers unfolding via multiple pathways in a rough energy landscape, where the timescale of interconversion between the folded states is assumed to be slow compared to that of unfolding. This scenario predicts that the nonexponential dwell times at a force of $110$pN are consistent with exponentially distributed free energy barriers~\cite{Brujic2006}. By contrast, a more recent work assumes that the static disorder~\cite{Zwanzig1990, Zwanzig1992} has a Gaussian distribution of barriers and derives the corresponding function to fit the experimental dwell times over a range of constant forces~\cite{Kuo2010}. Alternatively, assuming that the Gaussian distribution comes from the noise in the applied force~\cite{Brujic2007}  leads to the same form of the non-exponential fitting function for the dwell times if the noise correlation time is longer than that of unfolding. In addition to these physical interpretations, in~\cite{Garcia-Manyes2007} a log normal distribution is proposed to be the best heuristic fit to the dwell times of both monomeric and polyubiquitin data.

A possible explanation for the apparent success of these four models in fitting the same data is that rigorous methods of analyzing and assessing force-clamp trajectories are lacking. For example, some studies average and normalize the measured end-to-end length trajectories as an estimate of the cumulative unfolding probability, while others export the individual dwell times and bin them into probability density distributions before fitting. Moreover, since the polyprotein chains vary in length and detach from the cantilever at random times, not all events are necessarily observed in the experiment. In order to account for the undetected events different filtering methods are applied to the data, each with their own associated uncertainties. In this paper we quantitatively assess the errors in existing analysis protocols, develop new analysis methods that systematically take into account biases introduced by experimental artifacts and evaluate the success of each model using not only graphical tests, but also rigorous statistical tests based on maximum likelihood estimation and Bayesian sampling. We show that tests of robustness against filtering the data provide an excellent indication of the validity of the underlying model and we illustrate the results in both real and synthetic data sets. In order to use the full experimental data set and avoid filtering, we additionally derive a likelihood function that calculates the probability of observing a sequence of dwell times followed by the measured detachment time of the molecule. This method allows us to rank the proposed models in terms of their consistency with observing the data set using standard statistical tests, such as the Kuiper test~\cite{Kuiper1960, Tygert2010}. Finally, we show the agreement between filtering techniques and the use of the likelihood function and propose a self-consistent recipe for data assessment in future experiments.

The importance of distinguishing between fitting functions is to deduce the correct physical picture for protein unfolding, which sets the mechanical response timescales in biology. Indeed, it is striking that the mean unfolding times for the four proposed distributions for ubiquitin span from seconds to hundreds of seconds at a given constant force, thus emphasizing the importance of determining the correct model.

\section*{Materials and Methods}

\begin{figure}
\begin{center}
\includegraphics[width=5.25in]{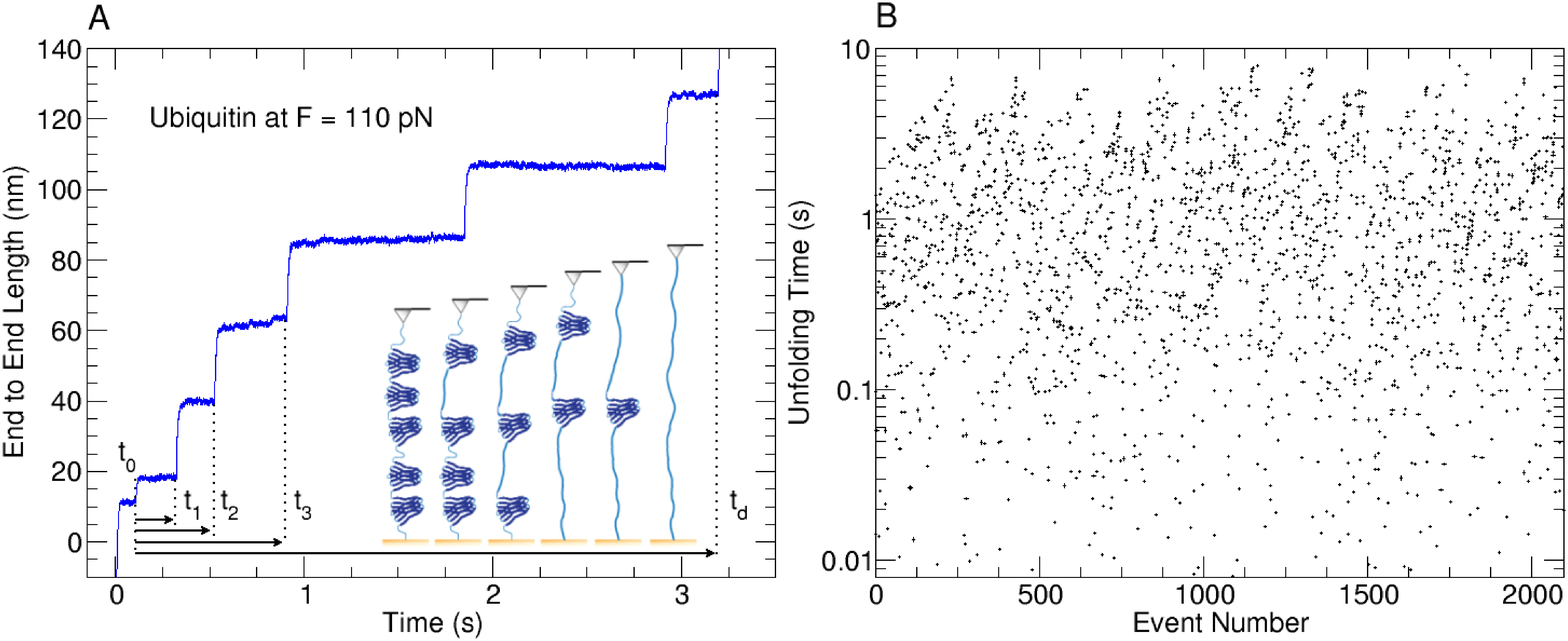}
\caption{}
\label{fig:one}
\end{center}
\end{figure}

Force-clamp spectroscopy measurements are taken using the same AFM, ubiquitin polyprotein construct and experimental method described in~\cite{Oberhauser2001, Fernandez2004, Schlierf2004, Brujic2006, Brujic2007, Garcia-Manyes2007, Garcia-Manyes2009, Kuo2010}. In response to a constant stretching force, each of the protein domains in a polyprotein chain unfolds stochastically, leading to a stepwise elongation of the end-to-end length over time, as shown in the example in Fig.~\ref{fig:one}~A. Time zero is marked at the beginning of the first plateau in the end-to-end length after the constant stretching force of~110pN is applied. The resulting staircase of unfolding events yields a set of dwell times $t_1$, $t_2$, ..., which mark the rupture of the native state of each domain to the fully extended unfolded state. Only staircases with a minimum of $3$ repeating steps are included in the analysis as the signature of the single polyprotein molecule. Plotting over~2000 unfolding times in the order in which they are collected leads to the scatter graph in Fig.~\ref{fig:one}~B. The logarithmic scale emphasizes the span over three orders of magnitude of the unfolding times, while the homogeneity of the data from experiments performed with distinct cantilevers and on different days gives validity to the force calibration and the stability of the protein, respectively.    

\section*{Results and Discussion}

\subsection*{Unbiasing the unfolding data from experimental artifacts}

In order to determine the probability $F(t)$ of observing an unfolding event after a time $t$, a common way to analyze time series data is to plot the cumulative distribution function (CDF) of the dwell times. Experimentally this CDF is often constructed by averaging and normalizing the raw staircases, but this method gives an approximation of the CDF that is not monotonically increasing due to the presence of thermal noise and occasional drift in the experiment. Instead, the correct way to construct the CDF is to directly export the dwell times, sort and rank them from smallest to largest and then plot the normalized rank against the dwell time as the empirical CDF. This procedure avoids loss of information by binning, given that this empirical CDF has a value at each measured dwell time.

However, in the case of force-clamp trajectories the empirical CDF of all the observed dwell times does not coincide with the unfolding probability $F(t)$ because of experimental artifacts. The experimental window (fixed by the time resolution at short times, $\tmin$, and the total duration of the experiment, $\tmax$) may not encompass the whole range of the unfolding probability $F(t)$. The empirical CDF, given by
\begin{equation}
  \label{eq:5}
  \hat P(t) = \frac{\#\{\text{dwell times} < t\}}{N}
\end{equation}
where $N$ is the total number of dwell times in the data set and $\#\{\text{dwell times }< t\}$ denotes the number of such times that are less than $t$, must therefore be fit with a $P(t)$, conditional on the time range of the experiment~\cite{Koster2006}. While $F(t)$ is zero at time zero and reaches one at infinity, the conditional $P(t)$ is fixed to zero at $\tmin$ in our experiments, reaches one at $\tmax$ and is defined as
\begin{equation}
  \label{eq:fitting}
   P(t)=
   \begin{cases}0 & \qquad \text{if\ $t < \tmin$}\\
     \displaystyle\frac{F(t)-F(\tmin)}{F(\tmax)-F(\tmin)} & \qquad \text{if\ $\tmin\le t\le \tmax$}\\
     1 & \qquad \text{if\ $t > \tmax$} \end{cases}
\end{equation}
Note that this conditional fitting of the data fixes the values of $F(t)$ at $\tmin$ and $\tmax$ without the need of introducing extra parameters. Some previous studies introduce a normalization constant as an extra fitting parameter $A$, which may unjustly improve the proposed fit to the theory. The functional form of $F(t)$ chosen for the fitting procedure self-consistently determines the range captured by the data, as shown in Fig.~\ref{fig:two}~A. If the experiment lasts long enough that the value of $F(t)$ approaches one at $\tmax$ then the conditioning has little effect on the parameters. However, even cases where $F(t)$ reaches $0.85$ at $\tmax$ can alter the rate of an exponential function by $27\%$ and change the shape of the distribution unless this conditioning is taken into account (see Fig. S1 in the Supplementary Materials).

Another artifact of force-clamp trajectories is that the molecules detach from the cantilever at random times $t_d$, which implies that some events are not observed in the experiment. If the total number of domains $N$ in the polyprotein chain were known \textit{a~priori}~\cite{YiCao2008}, one could unbias the distribution of dwell times using order statistics, assuming that the unfolding events are independent of one another~\cite{Brujic2007}. However, in our experiments the cantilever picks up polyproteins at random points on the surface such that any $N$ (up to the full length $N_\text{max}$) can be exposed to a stretching force in a given experiment. This renders the unbiasing procedure difficult to resolve as different distributions $p(N)$ bias the empirical $\hat P(t)$, which is illustrated on synthetic examples in Fig. S2 in the Supplementary Materials. It is therefore necessary to filter the data, such that all events come from trajectories that correspond to the same time window in the experiment, in order to construct the $P(t)$ that corresponds to the underlying $F(t)$. The correct way to do so is to choose an experimental time window (e.g. from $\tmin$ to a cutting time $t_{c}$) and only consider those dwell times that (i) occur within that range and (ii) come from trajectories that lasted over the entire range, such that $t_d\ge t_{c}$, as shown in Fig. S3 in the Supplementary Materials.
Note that filtering the data by the detachment time alone by keeping all dwell times less than $t_d$ leads to empirical CDFs that give inaccurate values of the fitting parameters, as shown in Fig. S4 in the Supplementary Materials. 

\subsection*{Graphical tests of the unfolding probability $F(t)$}
\label{norm}

\begin{figure}
\begin{center}
\includegraphics[width=5.25in]{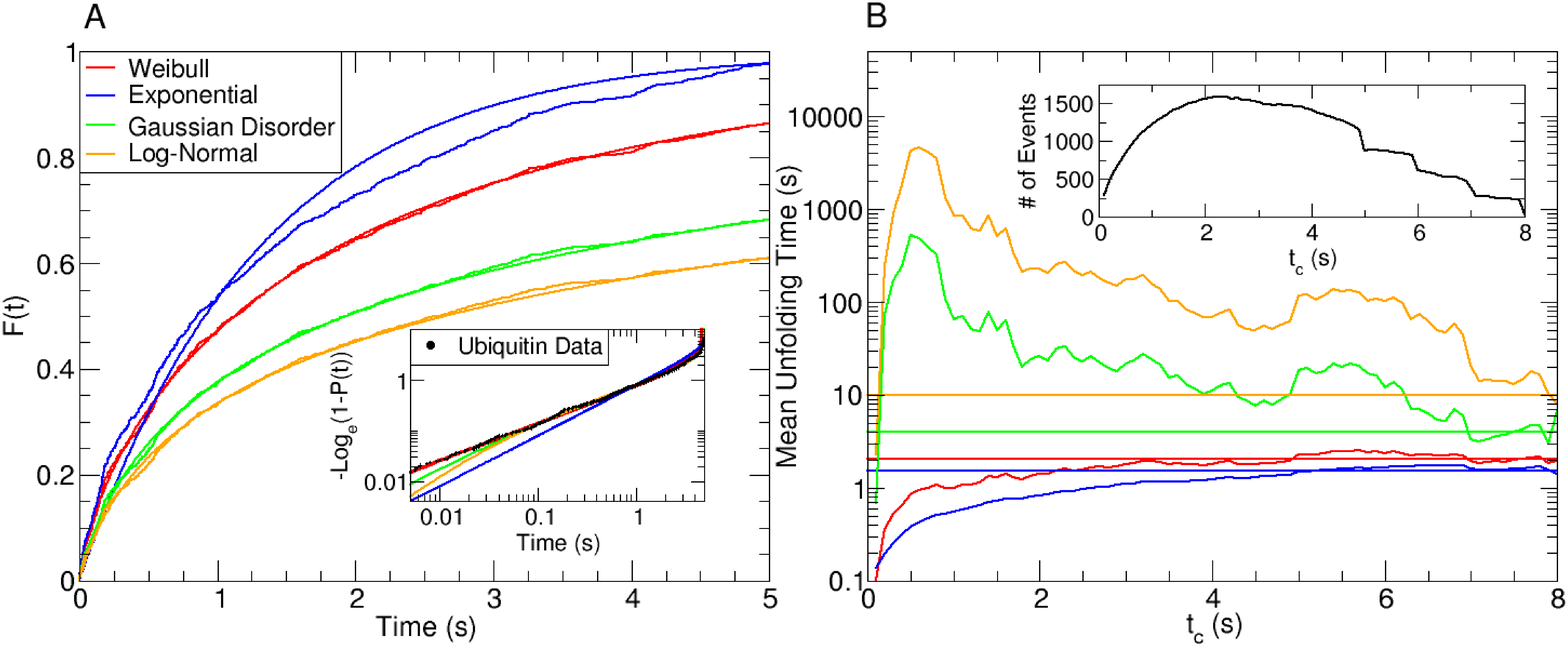}
\caption{}
\label{fig:two}
\end{center}
\end{figure}

Using the described methods for filtering and fitting of the experimental CDF $\hat P(t)$, we assess the success of different models in explaining the ubiquitin data. The experimental time window is chosen to be between the time resolution of the experiment $\tmin=5$ms and the cutting time $t_{c}=5$s, which ensures three decades over which to test the goodness of fit of the data. The same empirical $\hat P(t)$ is then fit with Eq.~\eqref{eq:fitting} for the four functional forms of $F(t)$ proposed in the literature and listed in Table 1. The fitting can be done by least squares or maximum likelihood methods, which result in parameters that agree to within two decimal places. Since the fitting procedure self-consistently fixes $F(\tmax)$ and $F(\tmin)$ for each function, the resulting empirical $\hat F(t)=(F(\tmax)-F(\tmin)) \hat P(t) +F(\tmin) $, obtained by solving Eq.~\eqref{eq:fitting} for $F(t)$, differ in their range, as shown in Fig.~\ref{fig:two}~A. For instance, the experimental window captures only $60\%$ of the events in the case of the log normal distribution, while it covers almost all the events in the case of the exponential function. Moreover, the curves clearly show that the exponential fitting is inaccurate, while the other three models are all in good agreement with the data on the linear scale and exhibit comparable $\chi^2$ values.  In order to zoom into the two decades of fast unfolding times, the inset shows the data plotted as the conditional $P(t)$ on a log-log scale that emphasizes deviations from the fits. Here it can be seen that the Weibull distribution performs better than all others on timescales below $0.1$ seconds. Note that the Weibull distribution plotted as the $F(t)$ would be a straight line on the scales of the inset, but the $P(t)$ distribution is conditional on the time window of the experiment and thus exhibits curvature. Even though the Weibull distribution fits this data set most accurately, the statistical error in the experiment precludes the determination of the correct model by this graphical test alone.

Indeed, many functional forms (particularly those with several parameters) can be successful in fitting a particular time window chosen for the analysis, but it is a greater challenge to assess how robust the fitting function and its parameters are against filtering the same data over different time windows. The shorter the time window, the more data points are needed to obtain the same statistical accuracy in the fitting, as shown in the synthetic example in Fig. S5 in the Supplementary Materials. Nevertheless, there exists a range of cutting times $t_c$ over which the fitted parameters should converge to the same values given a large enough pool of data. As a test of robustness of the parameters, we calculate the first moment of $F(t)$ (i.e. the mean unfolding time) fitted at different values of $t_c$, shown in Fig.~\ref{fig:two}~B. It can be seen that filtering the data at any time above $2.5$s has little effect on the mean unfolding time for the Weibull distribution, while the Gaussian disorder and log normal distributions vary greatly with $t_c$. The mean unfolding time is plotted on a logarithmic scale in order to capture the three orders of magnitude span that is predicted by the different experimental time windows of the same pool of data. This result shows that fitting with different physical models leads to dramatic consequences on biological function, since the characteristic protein unfolding time varies from $1$ second to $3$ minutes.

While it can be argued that the statistical pool of filtered data shown in the inset is insufficient to fit $F(t)$ at short cutting times $t_c$, the lack of convergence over any significant filtering range for the Gaussian disorder and log normal functions questions their validity in describing the data. On the other hand, the exponential distribution does exhibit a range of stability after $t_c\approx3$s, but its poor performance in fitting the data invalidates its use for a different reason. This analysis shows that a successful model must not only fit the data with fidelity over a range of $t_c$, but also predict parameters that are stable over that range. 

\begin{figure}
\begin{center}
\includegraphics[width=5.25in]{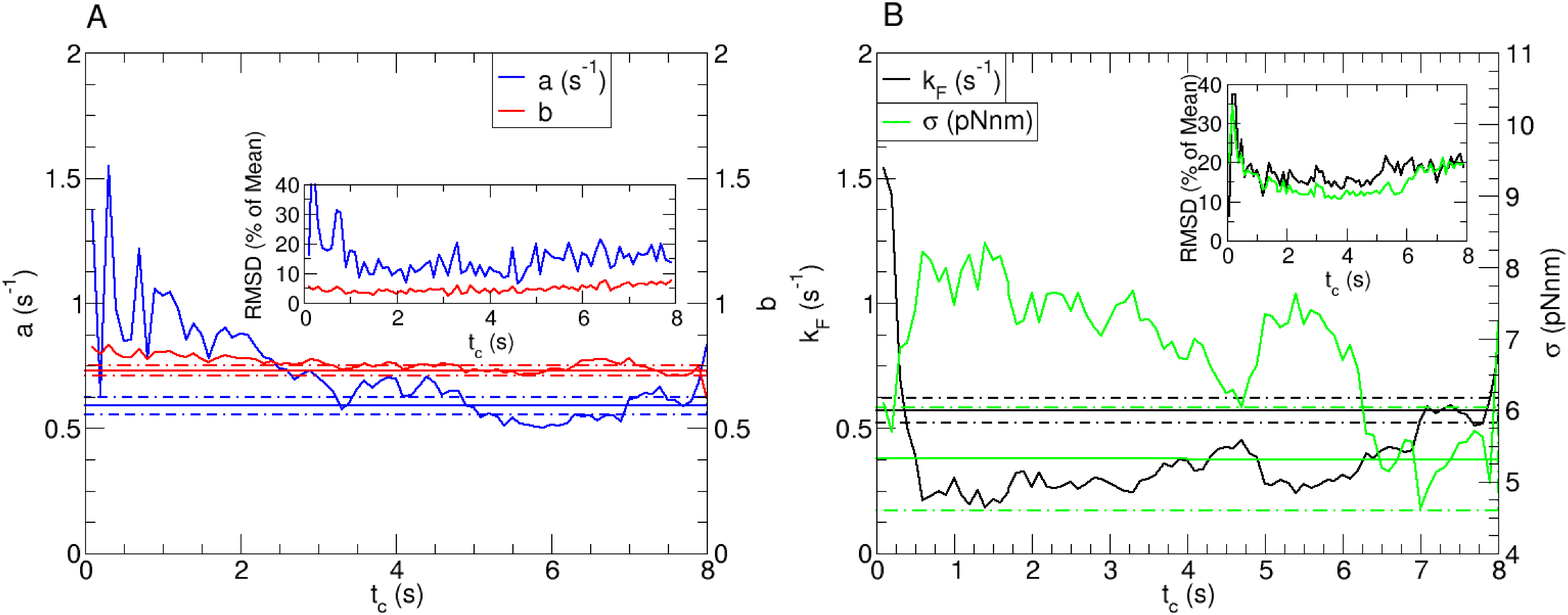}
\caption{}
\label{fig:three}
\end{center}
\end{figure}

Instead of using the least squares method to assess the goodness of fit and extrapolate variance in the parameters by bootstrapping, other approaches work equally well. One such method is Maximum Likelihood Estimation (MLE)~\cite{Edwards1992}, which computes the most likely parameters of a distribution using a set of variables - in our case the dwell times. The variance in the parameters is then obtained by Bayesian sampling of the data set~\cite{Howson2005}, as shown in Fig. S6 in the Supplementary Materials. The mean values of parameters $a$ and $b$ in the Weibull distribution and $k_f$ and $\sigma$ in the Gaussian disorder distribution are shown as a function of the experimental time window $t_c$ in Fig.~\ref{fig:three}~A,B, respectively. The inset shows that the root mean standard deviation (RMSD) in the fitting parameters decreases as a function of $t_c$, which is consistent with the concomitant increase in the number of data points and the wider time window of the fit. While the fluctuations observed in the Weibull parameters converge to stable values above $t_c\approx3$s, those of the Gaussian disorder model do not settle to any given values before $t_c\approx7$s, which is also reflected in the broad fluctuations of the mean unfolding time shown in Fig.~\ref{fig:two}~B. Note that filtering at long $t_c$ uses as little as $10\%$ of the data collected, as shown in the inset. Disregarding the majority of the data set is never desirable to an experimentalist.

\subsection*{Maximum likelihood function includes all collected data}
\label{mle}

An alternative MLE function to fitting the CDF of the dwell times as a function of $t_c$ is one that takes into account experimental features of force-clamp trajectories and thus uses the whole data set to estimate parameters in the unfolding model. In a typical pulling experiment, the cantilever picks up a polyprotein chain of $N$ domains with a probability~$p(N)$. These domains subsequently unfold at dwell times $t_1$, $t_2$, \ldots, $t_k$, where $k$ corresponds to the last observed step in the staircase with a minimum $k_*=3$ required as the signature of the single molecule.  Finally, the molecule detaches either from the tip or the surface at time $t_d$. Assuming that the dwell times are independent of one another~\cite{Brujic2007,YiCao2008} and identically distributed~\cite{Bura2007, Bura2008, YiCao2011} we calculate the probability of observing $k$ unfolding events, multiplied by the probability of $N-k$ domains remaining folded up to the detachment time $t_d$, for every polyprotein chain:
\begin{equation}
  \label{eq:1}
  \frac{1}{G_*} \sum_{N=k}^{N_*} p(N)
  \frac{N!}{(N-k)!k!} f(t_1)\cdots f(t_k) [1-F(t_d)]^{N-k}
\end{equation}
where $f(t)= -dF/dt$ is the probability density associated with $F(t)$, $N_*=12$ is the number of domains in the expressed protein construct and $G_*$ accounts for the probability of not including staircases with less than $k_*=3$ steps
\begin{equation}
  G_*=\sum_{N=k_*}^{N_*}p(N)\sum_{l=k_*}^{N}\frac{N!}{(N-l)!l!}[F(t_d)]^l[1-F(t_d)]^{N-l}
\end{equation}
Taking the product of the likelihoods for each polyprotein chain in Eq.~(\ref{eq:1}) gives the overall likelihood function. The parameters in the unfolding probability $F(t)$ as well as those defining $p(N)$ (assumed to be a power law with a decay coefficient $\gamma$ in this case) are obtained by maximizing this likelihood function, while the uncertainties are estimated using Bayesian sampling. 

The maximum value of the likelihood function from the ubiquitin data set ranks the four proposed unfolding distributions in the following order from highest to lowest likelihoods: Weibull, Gaussian disorder, log normal and exponential distribution. Given that the actual values of the likelihoods depend on the size of the data set and $F(t)$, this rank test only estimates which distribution is more consistent with the data, but it cannot assess the accuracy of the fits themselves. Nevertheless, the fact that the Weibull parameters from the likelihood function, also shown in Fig.~\ref{fig:three}~A, are in good agreement with the parameter convergence of the fits of the $F(t)$ above $t_c\approx3$s gives further support to this model. Conversely, the lack of such an agreement in the runner-up Gaussian disorder model suggests that this is not the correct functional form for the unfolding probability. 

\begin{figure}
\begin{center}
\includegraphics[width=5.25in]{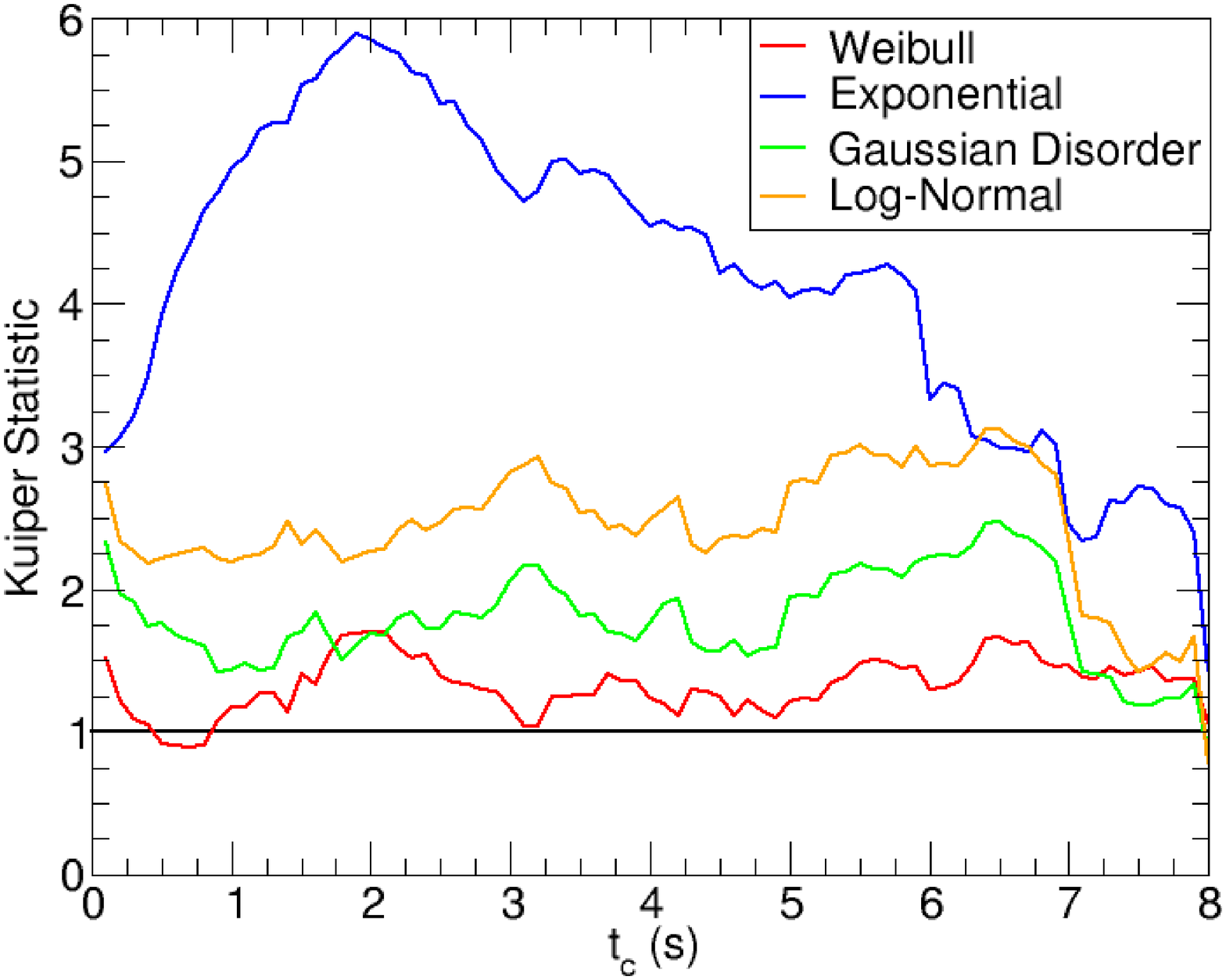}
\caption{}
\label{fig:four}
\end{center}
\end{figure}

A statistical test that quantitatively assesses whether a set of observables originates from a given distribution is the Kolmogorov-Smirnov approach with a modification by Kuiper~\cite{Kuiper1960}. We therefore compare the empirical CDFs at different $t_c$ from the data set with those generated from the four functional forms of $f(t)$ using the parameters estimated by the above likelihood function. Denoting as before by $P(t)$ the postulated distribution and by $\hat{P}(t)$ the experimental distribution, the Kuiper statistic is defined as, 
\begin{equation}
  U = \sqrt{N}\max_{j=1,\ldots,
    N}\left(P(t_j)-\hat{P}(t_j)\right)-\sqrt{N}\min_{j=1,\ldots, N}\left(P(t_j)-\hat{P}(t_j)\right)
\end{equation}
where the maximum and the minimum are taken over all the $N$ dwell times $t_1, \ldots, t_N$ in the data set. $U=1$ signifies a perfect match. The results in Fig.~\ref{fig:four} show that the Weibull distribution is closest to $1$ over almost the entire range of $t_c$. While the Gaussian disorder model is slightly closer to $1$ between $7.2<t_c<8$s, this narrow range is based on less than $10\%$ of the collected data. 

\subsection*{Comparison with synthetic and other data sets}

\begin{figure}
\begin{center}
\includegraphics[width=5.25in]{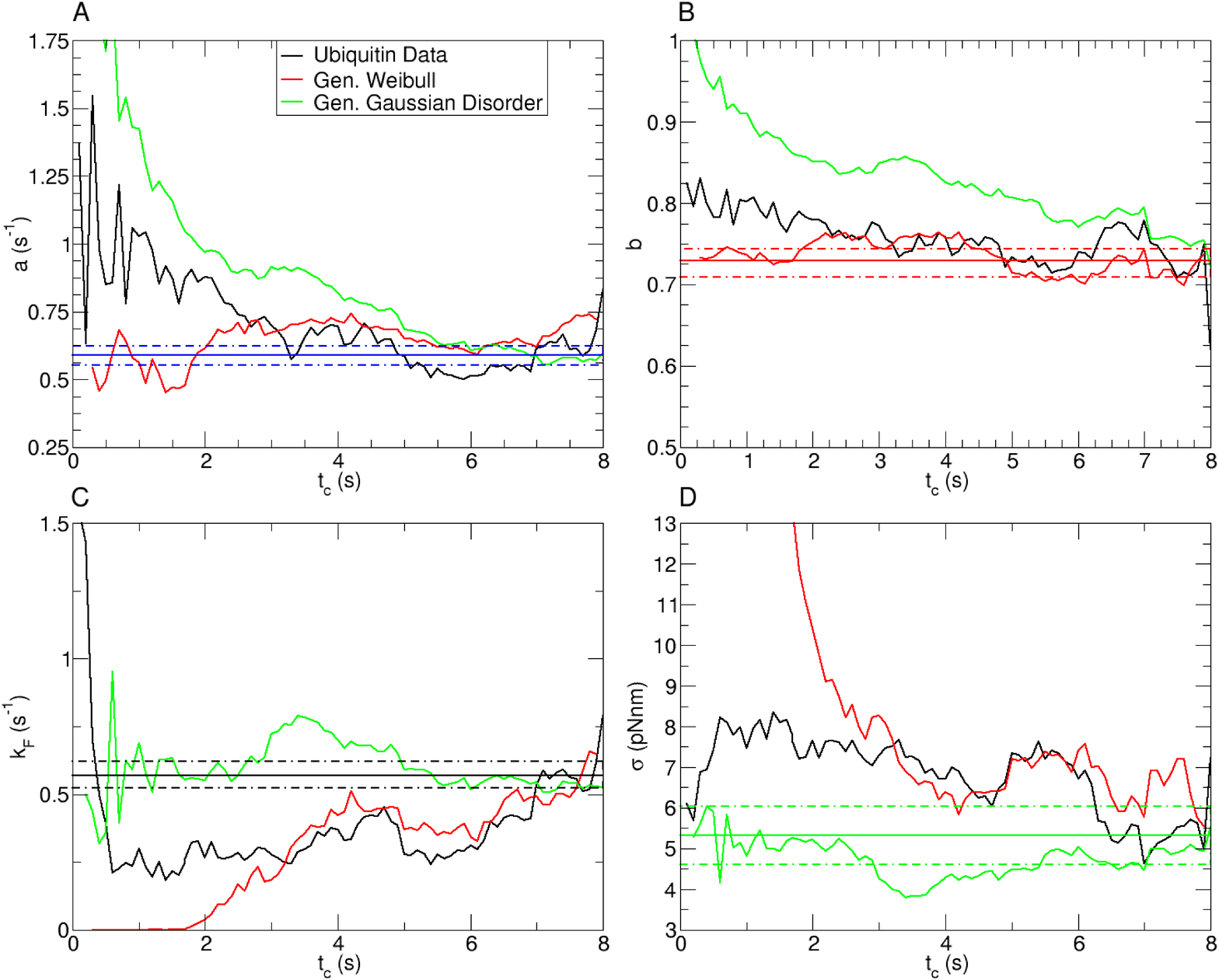}
\caption{}
\label{fig:five}
\end{center}
\end{figure}

In order to further test the consistency of the ubiquitin data with the Weibull and Gaussian disorder models, we generate two synthetic data sets using the parameters obtained from the maximum likelihood function in Eq.~\eqref{eq:1} that mimic the size of the experimental data. We then filter the synthetic and experimental data sets by $t_c$ and compare the values of their fitting parameters using the Weibull distribution in Fig.~\ref{fig:five}~A,B and the Gaussian disorder model in Fig.~\ref{fig:five}~C,D. In all cases, we find that the experimental and the synthetic Weibull data are in good agreement with each other above $t_c=3$s, while the synthetic Gaussian disorder data exhibits significant deviations. The two data sets are similar in that they not only exhibit comparable fluctuations in the fitting parameters of the Weibull arising from statistical errors, but they also follow similar trends in their discrepancy from the fitting parameters of the Gaussian disorder model. All these results are consistent with the hypothesis that the unfolding of ubiquitin data at 110~pN is most likely to originate from a Weibull distribution. 

The fact that the Gaussian disorder model does not agree with the data contradicts theories of static disorder~\cite{Kuo2010} and force noise~\cite{Clusel2011}, since they imply the same fitting function. While the former places the Gaussian noise in the barriers to unfolding, the latter does so in the constant force applied by the cantilever. Given that $\sigma=\sigma_F \Delta x$, where $\sigma_F$ is the noise in the applied force and $\Delta~x=0.23$~nm is the distance to the transition state, $\sigma$ in the barriers obtained from the MLE function translates to $21\%$ error in the force calibration, while the most stable value of the parameter against filtering gives $32\%$, as shown in Table 1. If this functional form had fit the data well, the estimated error in the force calibration is much higher than the measured error of $\approx5\%$~\cite{Ohler2007} and would thus give validity to the scenario of static disorder in the ubiquitin free energy landscape rather than that of the force noise.    

It is worth noting that there are several reasons for which our results are not in agreement with those published in the literature and why they also disagree between each other. First, a common mistake in fitting force-clamp data is to introduce a normalization constant as an extra fitting parameter. Instead, care must be taken to fit the conditional unfolding probability distribution over the experimental window with $P(t)$ and obtain $F(t)$ using Eq.~\eqref{eq:fitting}. Second, binning the distribution of unfolding times should be avoided as it effectively introduces an extra parameter into the fitting and loses resolution at short unfolding times. Third, filtering the data by accepting those trajectories that last a set minimum detachment time $t_d$ and including events that occur after that $t_d$ into the $P(t)$ biases the resulting distribution at long times, which in turn skews the fitting parameters. Instead, one must only include those dwell times that occur within exactly the same time window. Finally, plotting and fitting data on log-log scales can be useful if the data has been shown to fit well with a stretched exponential function in order to use a straight line fit. Otherwise, the compression of the data may obscure deviations from view and requires further assessment of the fits using MLE and Bayesian sampling. 
     
\section*{Conclusion}

Numerous force-clamp analysis methods, such as the fitting of filtered cumulative dwell time distributions, convergence of fitting parameters with an expanding time window of the experiment, the prediction of the maximum likelihood function for the whole data set, the Kuiper test, as well as the comparison with synthetically generated data sets, ubiquitously demonstrate that the data are most likely to arise from an underlying Weibull distribution, otherwise known as the stretched exponential distribution. This type of kinetics has been observed in the case of DNA relaxation~\cite{Biancaniello2008}, thermally induced protein folding~\cite{Leeson2000, Chung2005}, protein binding~\cite{Hagen1995} and conformational dynamics in solution~\cite{Yang2003}. Microscopically, the stretched exponential has been attributed to multiple pathways in the protein landscape~\cite{Hagen1996} or memory effects~\cite{Kou2004}. Our results show that such complexities may also play a role in the protein's response to a pulling force at the single molecule level. 

One possible interpretation is that the unfolding events can occur via many (random) pathways, each with a different rate $\alpha$, and the distribution of unfolding times is obtained via superposition of the exponential decays in each of these pathways. For example, the stretched exponential corresponds to rates that are distributed according to the L\'evy distribution, since its probability is defined implicitly via
\begin{equation}
  \label{eq:2}
  \int_0^\infty (1-e^{-\alpha t}) \rho(\alpha) d\alpha = 1-e^{-(at)^b} \equiv F(t)
\end{equation}
where $\rho(\alpha)$ cannot be written in closed analytical form but it exhibits a power law $\propto\alpha^{-\gamma}$ at large $\alpha$. Therefore, the stretched exponential fitting function is in agreement with the theoretical model used in~\cite{Brujic2006} to fit the ubiquitin unfolding kinetics. By contrast, the Gaussian distribution of energies or force noise proposed in~\cite{Kuo2010} corresponds to a log-normal distribution in the rates in Eq.~\eqref{eq:2} via the Arrhenius assumption.

These methods invite previous studies to verify the accuracy of their results and provide a statistical toolbox for the analysis of future force-clamp studies. Moreover, it is possible to build on these techniques to take into account the particularities of a given experiment. For example, it is possible to introduce correlations between the domains within the likelihood function or assume a known $p(N)$ in the case of pre-pulled proteins. More generally, this type of analysis can be applied to other types of force-clamp measurements, such as the disulfide bond rupture kinetics~\cite{Wiita2006} or the disassociation of quaternary interactions between individual domains~\cite{Xu2012}.  

\section*{Acknowledgements} 
We would like to thank Jin Montclare for the expression of the ubiquitin polyprotein and Maxime Clusel for useful discussions. J.~B. holds a Career Award at the Scientific Interface from the Burroughs Wellcome Fund and was supported in part by New York University Materials Research Science and Engineering Center Award DMR-0820341 and a Career Award 0955621.

\bibliography{bibl}	 

\clearpage
\begin{table}[h]
\begin{center}
\begin{tabular}{|c|c|c|}
\hline
\textbf{Distribution $F(t)$} & \textbf{Previous Studies} & \textbf{MLE Parameters}\\ \hline
Exponential & $a\sim0.67$~s$^{-1}$~\cite{Garcia-Manyes2009} & $a=0.66\pm0.02$~s$^{-1}$\\ 
$1 - e^{-at}$ & & \\ \hline
Log-Normal & $\sigma=3.0$~\cite{Garcia-Manyes2007} & $\sigma=2.04\pm0.05$\\
$\frac{1}{2}\erfc[-\frac{\ln(t/t_0)}{\sigma \sqrt{2}}]$ & $t_0=0.005$~s & $t_0=1.26\pm1.08$~s\\ \hline
Gaussian Disorder (GD) & $k_F=0.73\pm0.03$~s$^{-1}$~\cite{Kuo2010} & $k_F=0.57\pm0.05$~s$^{-1}$\\
$1-\int_{\mathbb{R}} e^{-k_Fte^{-\beta r}} \frac{e^{-\frac{r^2}{2\sigma^2}}}{\sqrt{2\pi\sigma^2}}dr$ & $\sigma=3.47\pm1.16$~pNnm & $\sigma=5.32\pm0.72$~pNnm\\ \hline
Force noise = GD & $\Delta x=0.23$~nm & $\Delta x=0.23$~nm\\
with $\sigma=\sigma_F \Delta x$ & $\sigma_F=15.09\pm5.04$~pN & $\sigma_F=23.13\pm3.13$~pN \\ \hline
Weibull & $a\sim0.9$~s$^{-1}$~\cite{Brujic2006} & $a=0.59\pm0.04$~s$^{-1}$\\
$1-e^{-(at)^b}$ & $b=\gamma-1=0.8$ & $b=0.73\pm0.02$\\ \hline
\end{tabular}
\caption{Table of parameter values from previous and our study for different distributions applied to a data set of ubiquitin
  pulled at $110pN$ of constant force.}
\end{center}
\end{table}

\clearpage
\section*{Figure Legends}

\subsubsection*{Figure 1}
(A) A typical force-clamp unfolding trajectory of a single ubiquitin polyprotein pulled with a constant stretching force of 110~pN. The beginning of the plateau that precedes the staircase of unfolding events marks time zero $t_0$ as the moment when the molecule is held taught under the applied force. The dwell times are then measured as the time interval between $t_0$ and each of the unfolding steps. Finally, the molecule detaches at $t_d$. The stepwise unfolding is illustrated in the schematic diagram. (B) Unfolding dwell times from the staircases are plotted on a semi-log scale in the order that they are collected and show a broad and homogeneous distribution of times.     
 
\subsubsection*{Figure 2}
(A) The unfolding probability $F(t)$ for four models proposed in the literature is used to fit the same empirical CDF of dwell times. The normalization of each $F(t)$ leads to different timescales on which the data unfold. The inset shows the corresponding conditional $P(t)$ on a log-log plot to emphasize the goodness of fit at short times. (B) Changing the time window from 5 seconds in (A) to $t_c$ shows the variability in the characteristic unfolding time between the different models. They span more than three orders of magnitude and only the Weibull and the exponential distribution settle to a given value. The inset shows how the number of data points changes as the time window is expanded.     

\subsubsection*{Figure 3}
Estimate of the fitting parameters in the Weibull in (A) and the Gaussian disorder distribution in (B) as a function of the experimental time window. Bayesian sampling shows that the fluctuations around the mean of the parameters diminish as the time window increases. The constant solid lines are the parameter values obtained from the maximum likelihood function in Eq.~\eqref{eq:1} and the dashed lines are their standard deviation. 
  
\subsubsection*{Figure 4}
A Kuiper statistic of 1 signifies a perfect match between the experimental data and the proposed distribution. Deviations from the line at 1 quantify the disagreement between the maximum likelihood function estimate for the four models and the experimental data set as a function of the experimental time window $t_c$.

\subsubsection*{Figure 5}
Comparison between synthetic data sets generated using the parameters in the maximum likelihood function for the Weibull and Gaussian disorder distribution and the experimental data set of ubiquitin. The constant solid lines are the parameter values obtained from the maximum likelihood function in Eq.~\eqref{eq:1} and the dashed lines are their standard deviation. Fitting the three data sets using the Weibull distribution gives the fluctuations in parameter $a$ in (A) and $b$ in (B) and using the Gaussian disorder distribution gives $k_F$ in (C) and $\sigma$ in (D). While the ubiquitin data and the synthetic Weibull distribution behave similarly above $t_c=3$s in all cases, the synthetic Gaussian distribution is significantly different.   

\clearpage

\clearpage

\clearpage

\clearpage

\clearpage

\end{document}